\documentclass[12pt,preprint]{aastex}
\usepackage{natbib}
\usepackage{graphicx}
\shorttitle{Testing the CGG model}
\begin{document}
\title{Testing the Cubic Galileon Gravity model by the Milky Way rotation curve and SPARC data}
\author{Man Ho Chan, Hon Ka Hui}
\affil{Department of Science and Environmental Studies, The Education University of Hong Kong, Hong Kong, China}
\email{chanmh@eduhk.hk}

\begin{abstract}
Recently, the Cubic Galileon Gravity (CGG) model has been suggested as an alternative gravity theory to General Relativity. The model consists of an extra field potential term which can serve as the `fifth-force'. In this article, we examine the possibility whether this extra force term can explain the missing mass problem in galaxies without the help of dark matter. By using the Milky Way rotation curve and the Spitzer Photomery \& Accurate Rotation Curves (SPARC) data, we show that this CGG model can satisfactorily explain the shapes of these rotation curves without dark matter. The CGG model can be regarded as a new alternative theory to challenge the existing dark matter paradigm. 
\end{abstract}

\keywords{dark matter}

\section{Introduction}
Recent observations indicate some deviation of the observed galactic rotational curves from the curves predicted by visible mass distribution under Kepler's laws. With the mass distribution extracted from the luminosity of spiral galaxies, a typical galactic rotation curve should have a decreasing trend moving away from the center of the galaxy under Kepler's laws, while observations in recent decades show that a flattened rotation curve is the common case \citep{Sofue}. Such discrepancy naturally leads to the conjecture on the existence of dark matter which cannot be probed from luminosity. Besides, observations of hot gas in galaxy clusters also reveal the existence of dark matter \citep{Chen}. It is commonly believed that dark matter consists of some unknown massive particles, such as sterile neutrinos \citep{Dodelson} or weakly interacting massive particles (WIMPs) \citep{Profumo}. However, recent experiments for dark matter search show negative results. No dark matter particle has been discovered for a wide range of mass and energy \citep{Tan,Akerib,Cooley}. The null result of direct detection might indicate that the assumption of the existence of particle dark matter is wrong. 

Besides the null detection problem, when we compare the rotational speed from the Newtonian gravity theory and the observations, one may reckon that the disagreement between the rotational speed values occurs mainly at large distance from the center of the galaxies \citep{Sofue}. Furthermore, the entire shapes of rotation curves of many galaxies trace their baryonic mass distributions. It is also surprising that the combination of the contributions of dark matter and baryons results in a nearly flat rotation curve. The transition from baryon to dark matter domination is very smooth. This problem is now known as the halo-disk conspiracy (for spiral galaxies) \citep{Battaner} or the dark-spheroid conspiracy (for elliptical galaxies) \citep{Remus}. 

Therefore, some suggest that the missing mass in galaxies and galaxy clusters can be explained by alternative theories of Newton's law or gravitational law so that no dark matter exists. The earliest version is the Modified Newtonian Dynamics (MOND) \citep{Milgrom}. Although this theory can explain some observations successfully, it works very poor in galaxy clusters \citep{Sanders,Sanders2}. Besides, some studies suggest that MOND is equivalent to some particular form of dark matter profile which suggests that the success of MOND is just a delusion \citep{Chan}. On the other hand, some suggest that the missing higher order terms in the current gravity theory makes itself only applicable to local space, including extra terms may explain why the galactic rotational curves usually flattens out. For instance, the Modified Gravity (MOG) theory suggests that some extra terms in the gravitational law can mimic the effect of dark matter \citep{Moffat}.  

In this article, we test a new alternative gravity theory proposed by \citet{Koyama,Sakstein} with the Milky Way rotation curve and the Spitzer Photomery \& Accurate Rotation Curves (SPARC) data \citep{Sofue2,Lelli}. We show that this model can also provide a viable solution to the missing mass problem. The Cubic Galileon Gravity (CGG) model includes an extra interaction term, which can be regarded as an additional force. By tuning the scaling parameter in the model, it provides a possible way to account for the significant boost in the galactic rotational speed at large galactocentric radius of the galaxies in our universe and the halo-disk (or dark-spheroid) conspiracy problem. Some previous studies have applied this model to galaxy clusters \citep{Salzano} and dwarf stars \citep{Sakstein}.

\section{The Cubic Galileon Gravity (CGG) Model}
There have been some new gravity theory alternatives to General Relativity arose among the community and one of them includes an additional field potential which results in the presence of a fifth-force as a supplement to the four fundamental forces \citep{Kimura,Koyama2}. The theories with a fifth-force are considered to possibly explain the anomalies of our universe observed in recent decades that do not agree with our existing theories. \citet{Sakstein} introduce an additional field $\phi$ to the Newtonian gravitational potential $\Phi_{\rm N}$ such that a fifth-force $F_5 = -\beta \nabla \phi$ comes into play, where $\beta$ is a dimensionless coupling parameter which plays an important role in determining the magnitude of the fifth-force.

Assume that the potential of the new field behaves the same way as the Newtonian potential, namely $\nabla^2 \phi = 8\pi \beta G \rho$. In order for the fifth-force to exhibit the Vainshtein Mechanism (screening effect at small galactocentric radius), a new derivative interaction term is added to the equation. By imposing spherical symmetry, the Poisson equation for the fifth-force potential field becomes \citep{Sakstein}
\begin{equation}
\frac{1}{r^2} \frac{d}{dr} \left( r^2 \frac{d \phi}{dr} \right) + \frac{1}{2\beta \Lambda} \frac{1}{r^2} \frac{d}{dr} \left[ r \left( \frac{d \phi}{dr} \right)^2 \right] = 8\pi \beta G \rho .
\end{equation}
where $\Lambda$ is a mass scale being of the order of the Hubble constant $H_0$. Integrating both sides and substituting the expressions of the fifth-force and the Newtonian gravitational force, one may obtain an equation for the ratio between the two forces $x = F_5/F_{\rm N}$ \citep{Sakstein}:
\begin{equation} \label{eq for x}
x+\left( \frac{r_{\rm V}}{r} \right)^3 \frac{x^2}{2\beta^2} = 2\beta^2, 
\end{equation}
where 
\begin{equation}
r_{\rm V}=r_{\rm V}(r) = \left[ \frac{GM(r)}{\Lambda^2} \right]^{1/3}
\end{equation}
is the Vainshtein Radius and generally has a dependence on $r$. One may realise that when the galactocentric radius $r$ is much greater than the Vainshtein Radius, $x \approx 2\beta^2$ and the fifth-force is unscreened. When $r$ is comparatively much smaller, the ratio $x$ (the strength of the fifth-force) is suppressed by a factor of $( r/r_{\rm V} ) ^{3/2}$. Such screening effect is called the Vainshtein Mechanism \citep{Sakstein}. The resultant force now has the magnitude of
\begin{equation} \label{total force}
F_{\rm total}=F_{\rm N}+F_5=(1+x)F_{\rm N},
\end{equation}
where the positive solution of $x$ is taken:
\begin{equation} \label{x}
x = \beta^2 \left( \frac{r}{r_V} \right)^3 \left[ \sqrt{1+\frac{4}{\beta} \left( \frac{r_{\rm V}}{r} \right) ^3}-1 \right]= \sqrt{\beta^4 k^6 + 4\beta^3 k^3}-k^3, 
\end{equation}
with $k= r/r_{\rm V}$.

\section{Consequence on the Galactic Rotational Speed}

The corresponding rotational speed is larger for a larger centripetal force. Within the screened regime, i.e.\ $r\ll r_{\rm V}$, the effect of the fifth-force is not significant enough to make the rotational speed deviate from that under the Newtonian gravitational force. However, in the unscreened regime, i.e. $r \gg r_{\rm V}$, the extra force term helps in boosting the rotational speed such that a flattened or even increasing rotational speed is possible at large galactocentric radius even without dark matter. The rotational speed can generally be given from the relation
\begin{equation} \label{v^2}
v^2(r) = rF_{\rm total} = r F_{\rm N}(r) (1+x) = (1+x) \frac{GM(r)}{r}.
\end{equation}
Note that $x$, the ratio between the fifth-force and the Newtonian force, is a function in $\beta$. Hence, adjusting the value of $\beta$ essentially changes the shape of rotation curve. In the CGG model, since $\Lambda$ is generally considered to be constant \citep{Sakstein}, $\beta$ is the only parameter that is adjustable. With a suitable value of $\beta$, the gravity model provides a possible candidate for the underlying functional form of gravity in order to account for the anomalous behavior of the galactic rotational curves.

\section{Curve-fitting for the Galactic Rotational Speed}

\subsection{Milky Way rotation curve}
First, we apply the CGG model to fit the Milky Way rotation curve data which can be obtained in \citet{Sofue2}. The baryonic components can be modeled by the SLFC model suggested in \citet{Flynn}. The SLFC model includes a potential to mimic the effect of dark matter in order to fit the data of the observed rotational speed. By taking away the potential term for dark matter, one may obtain the Newtonian prediction of the rotational speed for Milky Way using the visible mass profile in the SLFC model. Given the visible mass profile, one may also use the CGG model to boost the rotational speed. Our result shows that the CGG model can be an alternative theory of dark matter to describe the dynamics of Milky Way (see Fig.~1). Here, we assume $\Lambda=H_0=67.8$ km s$^{-1}$ Mpc$^{-1}$ \citep{Bucher}. The best-fit parameter is $\beta \approx 20$, which does not violate the observational constraint for our Solar system. As discussed in \citet{Sakstein}, the value of $r_{\rm V}$ for our Solar system is $\sim 100$ pc. Therefore, compared with the size of our Solar system ($\sim 100$ AU or $10^{-3}$ pc), the largest correction in the Newtonian gravitational force (see Eq.~(5)) within our Solar system is less than 1\% ($x \sim 10^{-6}$), which cannot be ruled out based on current observational data (for the Solar system constraints, see \citet{Will,Barreira}).

\begin{figure}
\vskip 10mm
 \includegraphics[width=140mm]{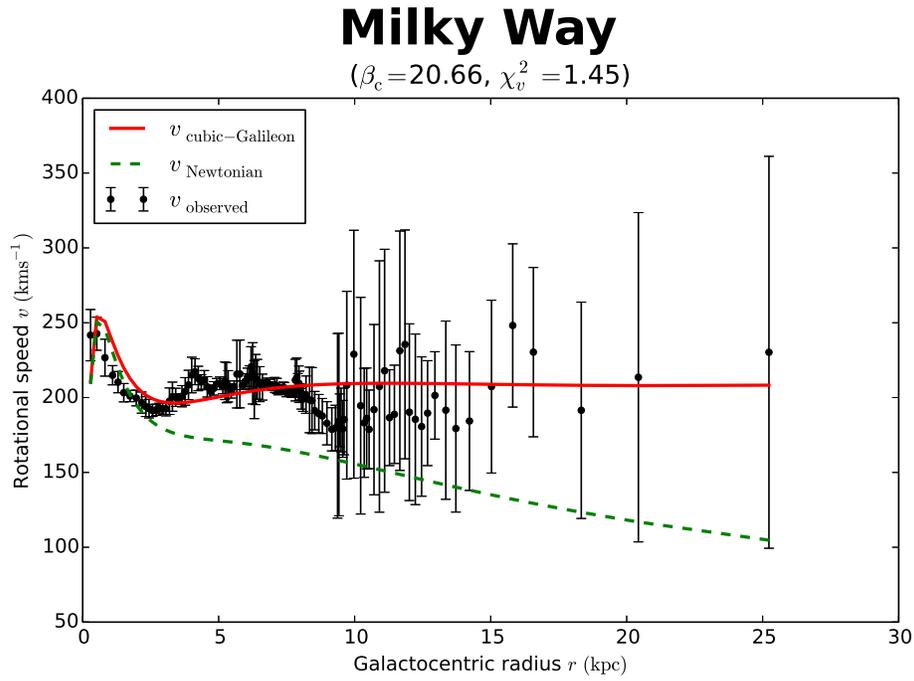}
 \caption{The resultant rotation curve of Milky Way. Here, the green dashed line is the rotation curve contributed by the baryonic component with $x=0$. The red solid line is the resultant rotation curve based on our model. The rotation curve data with error bars are taken from \citet{Sofue2}.}
\vskip 10mm
\end{figure}

\subsection{SPARC rotation curves}
Next, we apply the CGG model to other galaxies. We test our model with the SPARC (Spitzer Photometry \& Accurate Rotation Curves) obtained by \citet{Lelli}. Combining these data with the CGG model, we can give a further test for the possible range of $\beta$.

The baryonic mass profiles for different galaxies can be obtained by their corresponding luminosity profiles. In the database of SPARC, the luminosity contributed from the disk component and the bulge component are separated and presented as the velocity contributions $V_{\rm disk}$ and $V_{\rm bulge}$ correspondingly by assuming the mass-to-luminosity ratio to be the Solar ratio, i.e. $\Upsilon= M_\odot / L_\odot $. The mass inscribed up to galactocentric radius $r$ is given by
\begin{equation} \label{Mass}
M(r)= \frac{r}{G}V_{\rm effective}^2 = \frac{r}{G} \left[ \Upsilon_{\rm disk}V_{\rm disk} ^2 + \Upsilon_{\rm bulge}V_{\rm bulge} ^2  \right],
\end{equation}
where $\Upsilon_{\rm disk}$ and $\Upsilon_{\rm bulge}$ are the mass-to-luminosity ratios for the bulge and disk component respectively. Both $\Upsilon_{\rm disk}$ and $\Upsilon_{\rm bulge}$ are taken to be of the value $\sim 1$ respectively.

We are not going to use all data of SPARC (175 spiral galaxies) to test the CGG model. In particular, we choose the best candidates which satisfy the following 3 criteria: 1. Type S0-Sc (Hubble stage $T=0-5$) galaxy (i.e. ruled out Scd-Im); 2. Distance to the galaxy $D \le 50$ Mpc; 3. Small uncertainty of the distance to the galaxy ($\delta D/D \le 20$\% ). The use of the above criteria can be justified by the following arguments. Generally speaking, type Scd-Im (Hubble stage $T=6-10$) galaxies have `broken-arm' and diffuse features which made up of individual stellar clusters and nebulae \citep{Boeker}. Therefore, they usually have irregular shapes so that symmetry is difficult to apply in these galaxies. For distant galaxies ($D>50$ Mpc), the uncertainties in observations are usually quite large, especially in luminosity and rotational velocity determination. Lastly, there are some galaxies in the database of SPARC with large uncertainty in distance estimation (e.g. UGC 08699). The criteria $D \le 50$ Mpc and $\delta D/D \le 20$\% are reasonable cutoffs to minimize the uncertainties. Based on these reasons, we rule out galaxies which do not satisfy all of the above criteria 1-3 and finally have 28 galaxies for testing.

By using the mass profile in Eq.~(7), the rotational speed of the CGG model is calculated from Eq.~(6) and is compared with the observations. Obviously, different galaxies should have slightly different mass-to-luminosity values. Therefore, we have 3 parameters for fitting: $\beta$, $\Upsilon_{\rm disk}$ and $\Upsilon_{\rm bulge}$. For those galaxies without bulge, we set $\Upsilon_{\rm bulge}=0$. Similar adjustment of mass-to-luminosity ratio has also been found in MOND's framework \citep{Sanders2}. The best-fit values of the parameters are obtained when the reduced $\chi^2$ value is minimized. The reduced $\chi^2$ is defined as $\chi_v^2=(1/f)\sum_i(v_i-o_i)^2/s_i^2$, where $f$ is the degrees of freedom, $v_i$ are the calculated rotational speed in the CGG model, $o_i$ are the observed rotational speed and $s_i$ are the uncertainties of the observed rotational speed. We adapt the following criteria in tunning these three parameters: $\beta \sim 20$, $\Upsilon_{\rm disk} \sim 1$ and $\Upsilon_{\rm bulge} \sim 1$ \citep{Lelli}.

Generally speaking, the value of $\beta$ basically controls the magnitude of the rotational speed at large radius, lifting and lowering down the `tail' of the rotation curve, while the values of $\Upsilon_{\rm disk}$ and $\Upsilon_{\rm bulge}$ enlarge and diminish the contribution of the corresponding velocity component according to Eq.~(7). The relative values between $\Upsilon_{\rm disk}$ and $\Upsilon_{\rm bulge}$ also affect the shape of the graph. Galaxies generally have a bulge-dominant behavior when the value of  $\Upsilon_{\rm disk}$ is sufficiently larger than $\Upsilon_{\rm bulge}$, and vice versa. The best-fit results are shown in Figs.~2-8.

From the figures, we notice that the general shapes of the resultant rotation curves trace their baryonic distributions. This can provide an explanation for the halo-disk conspiracy problem. Also, the outer parts of the resultant rotation curves are generally flat which provide good agreement with the observational data. The best-fit values of $\beta$ fall within a small range $\beta \approx 12-27$ which suggests that $\beta$ is likely to be a universal constant. This is consistent with the CGG model. 

Note that some of the fits generate relatively large reduced $\chi^2$ values, especially for NGC5055. However, it does not mean that the CGG model works poor in these galaxies. In fact, the systematic uncertainties in obtaining the rotation curves are not completely negligible. For example, many rotation curve data in \citet{Lelli} are obtained by combining the H$\alpha$ data in the inner regions with HI data in the outer parts. It can give good quality of rotation curves because the H$\alpha$ data can trace the kinematics at high spatial resolutions so that the beam-smearing effects are minimal \citep{Lelli}. However, for NGC5055, its H$\alpha$ rotation curve has not been taken into account \citep{Lelli}. The resultant data points rely on HI data only, which may generate some systematic errors. Besides, the uncertainties due to the inclination of galaxies have not been considered either \citep{Lelli}. In particular, the uncertainty of the inclination of NGC5055 is greater than 10\%. Since the overall systematic error is relatively large while the observational uncertainty is small (less than 5\% for most of the data points) for those galaxies, the overall reduced $\chi^2$ values are somewhat overestimated. In fact, by comparing the rotation curves with the data points, we can see that the overall fits of rotation curves are not bad.

\begin{figure}
\vskip 10mm
 \includegraphics[width=140mm]{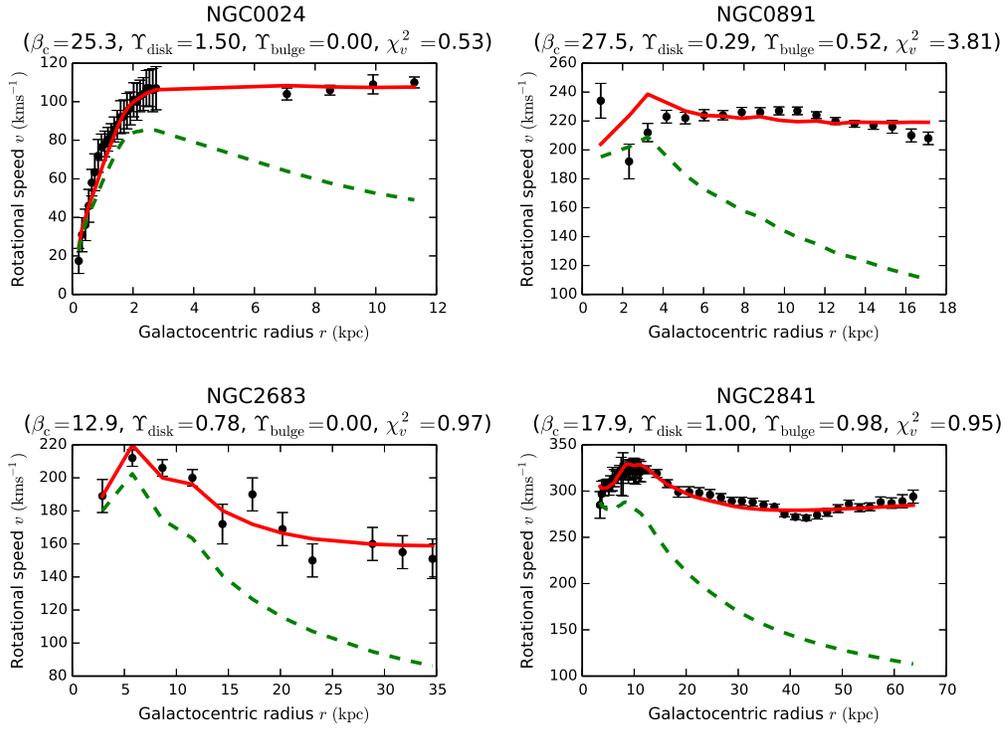}
 \caption{The resultant rotation curve of the NGC 0024, NGC 0891, NGC 2683 and NGC 2841 galaxies. Here, the green dashed line is the rotation curve contributed by the baryonic component with $x=0$. The red solid line is the resultant rotation curve based on our model. The rotation curve data with error bars are taken from \citet{Lelli}.}
\vskip 10mm
\end{figure}

\begin{figure}
\vskip 10mm
 \includegraphics[width=140mm]{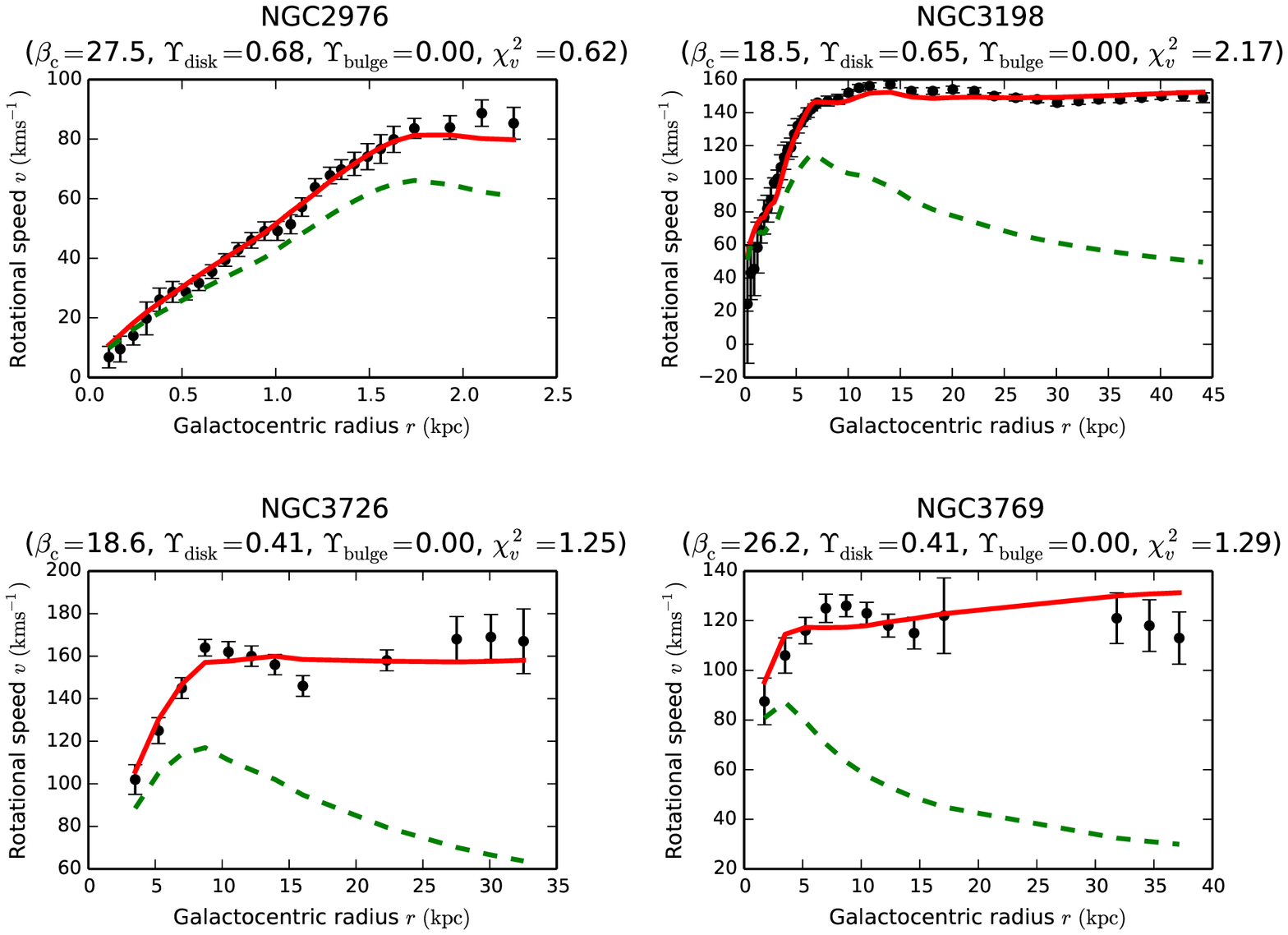}
 \caption{The resultant rotation curve of the NGC 2976, NGC 3198, NGC 3726 and NGC 3769 galaxies. Here, the green dashed line is the rotation curve contributed by the baryonic component with $x=0$. The red solid line is the resultant rotation curve based on our model. The rotation curve data with error bars are taken from \citet{Lelli}.}
\vskip 10mm
\end{figure}

\begin{figure}
\vskip 10mm
 \includegraphics[width=140mm]{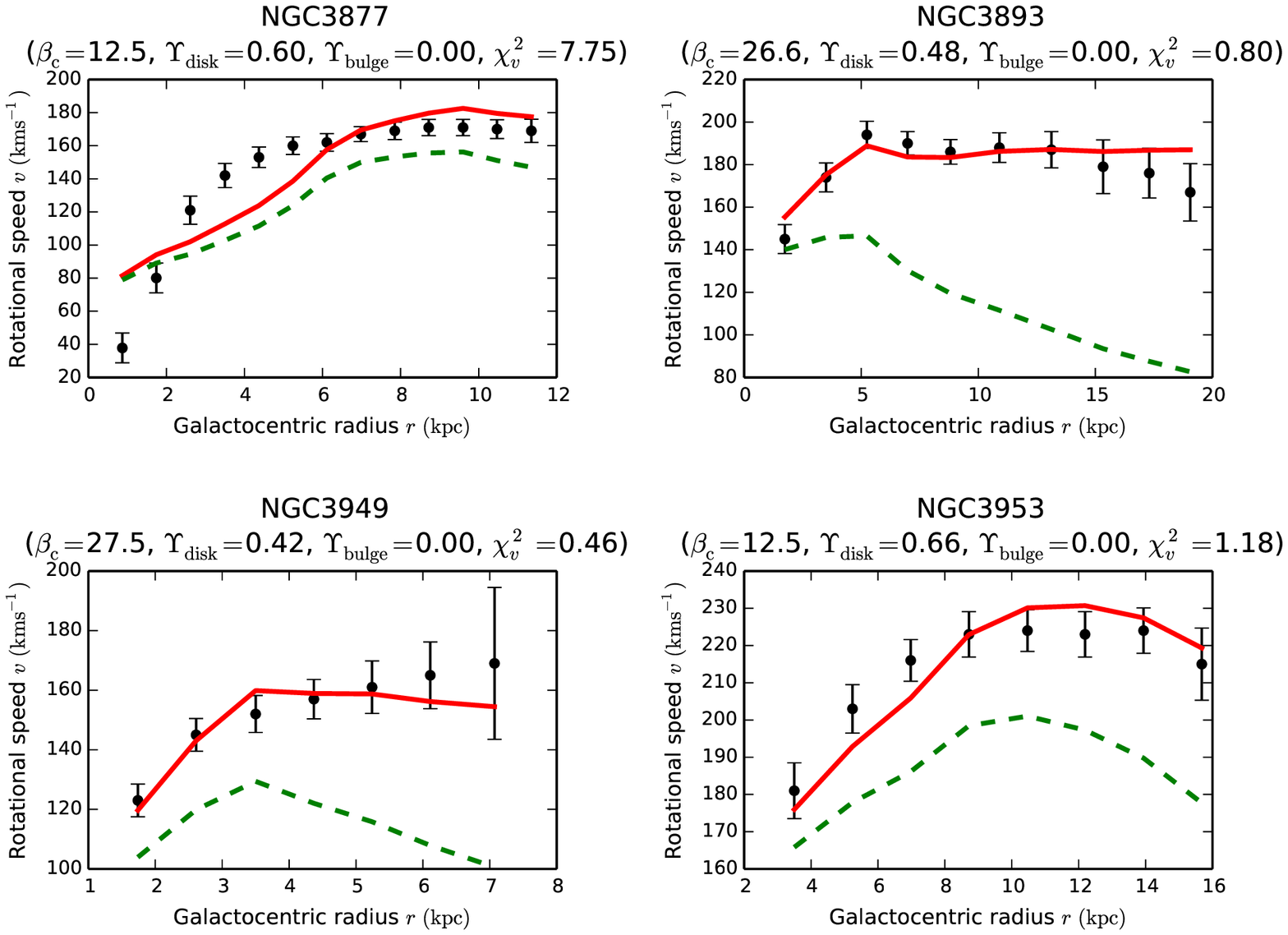}
 \caption{The resultant rotation curve of the NGC 3877, NGC 3893, NGC 3949 and NGC 3953 galaxies. Here, the green dashed line is the rotation curve contributed by the baryonic component with $x=0$. The red solid line is the resultant rotation curve based on our model. The rotation curve data with error bars are taken from \citet{Lelli}.}
\vskip 10mm
\end{figure}

\begin{figure}
\vskip 10mm
 \includegraphics[width=140mm]{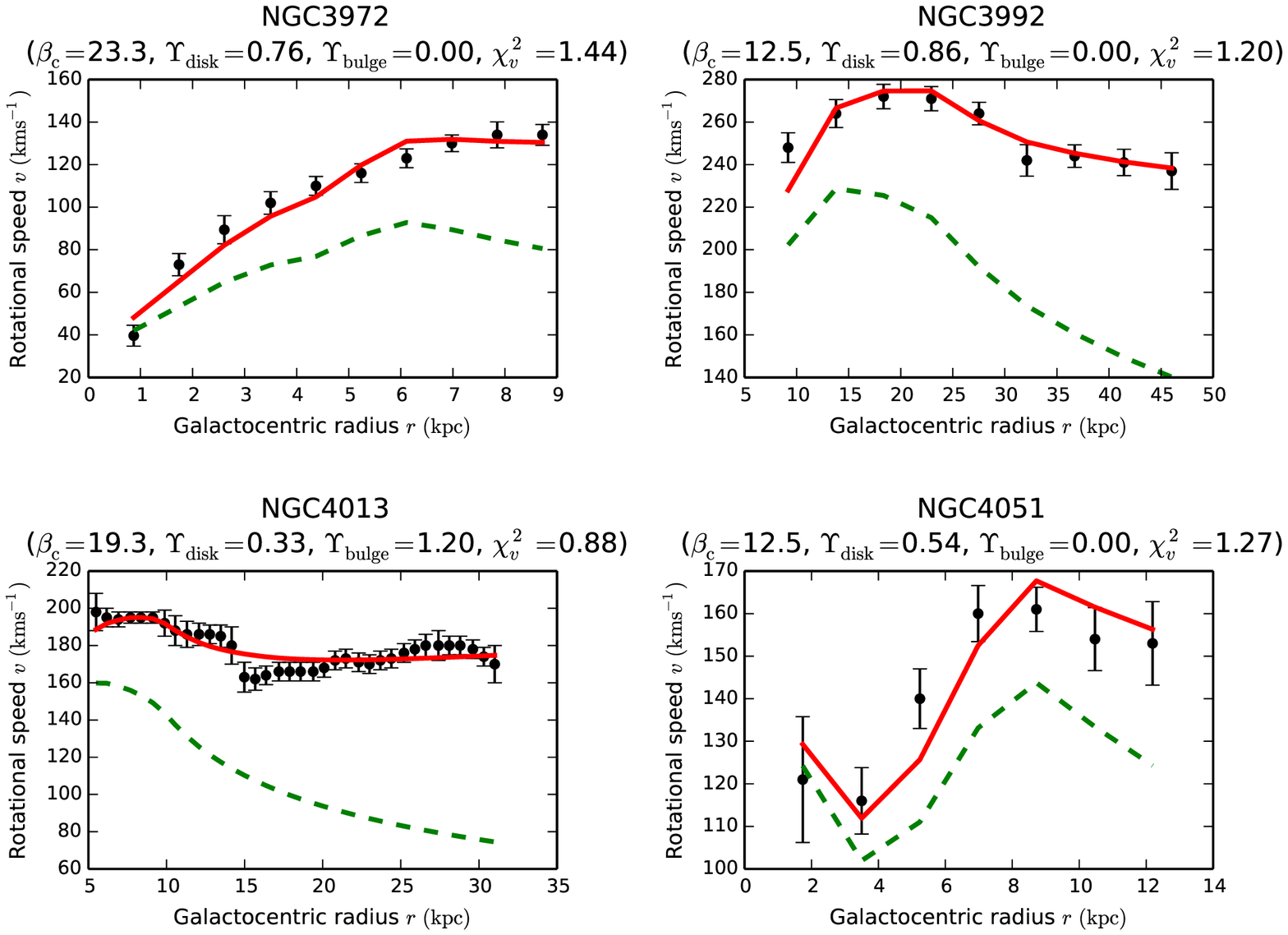}
 \caption{The resultant rotation curve of the NGC 3972, NGC 3992, NGC 4013 and NGC 4051 galaxies. Here, the green dashed line is the rotation curve contributed by the baryonic component with $x=0$. The red solid line is the resultant rotation curve based on our model. The rotation curve data with error bars are taken from \citet{Lelli}.}
\vskip 10mm
\end{figure}

\begin{figure}
\vskip 10mm
 \includegraphics[width=140mm]{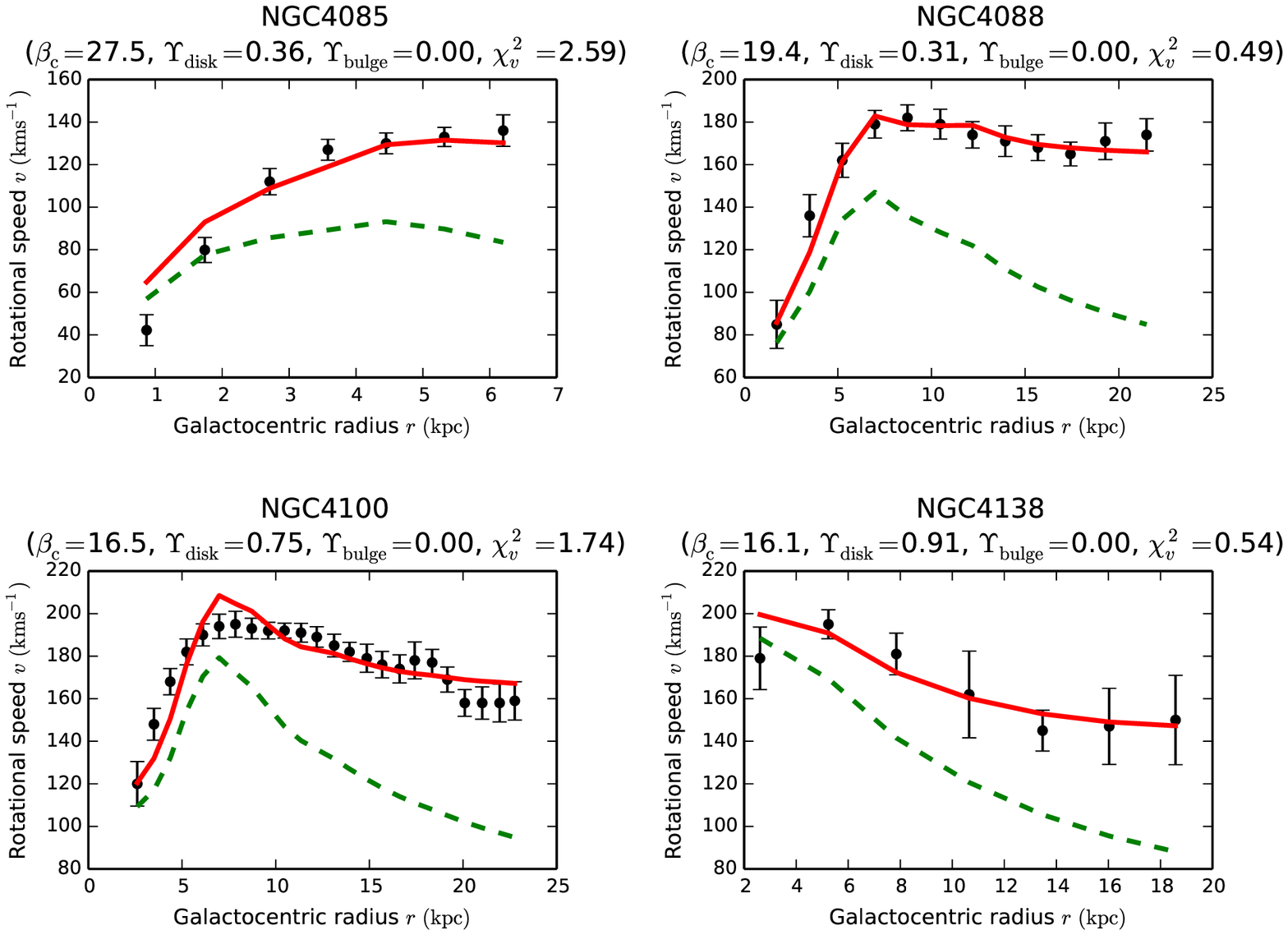}
 \caption{The resultant rotation curve of the NGC 4085, NGC 4088, NGC 4100 and NGC 4138 galaxies. Here, the green dashed line is the rotation curve contributed by the baryonic component with $x=0$. The red solid line is the resultant rotation curve based on our model. The rotation curve data with error bars are taken from \citet{Lelli}.}
\vskip 10mm
\end{figure}

\begin{figure}
\vskip 10mm
 \includegraphics[width=140mm]{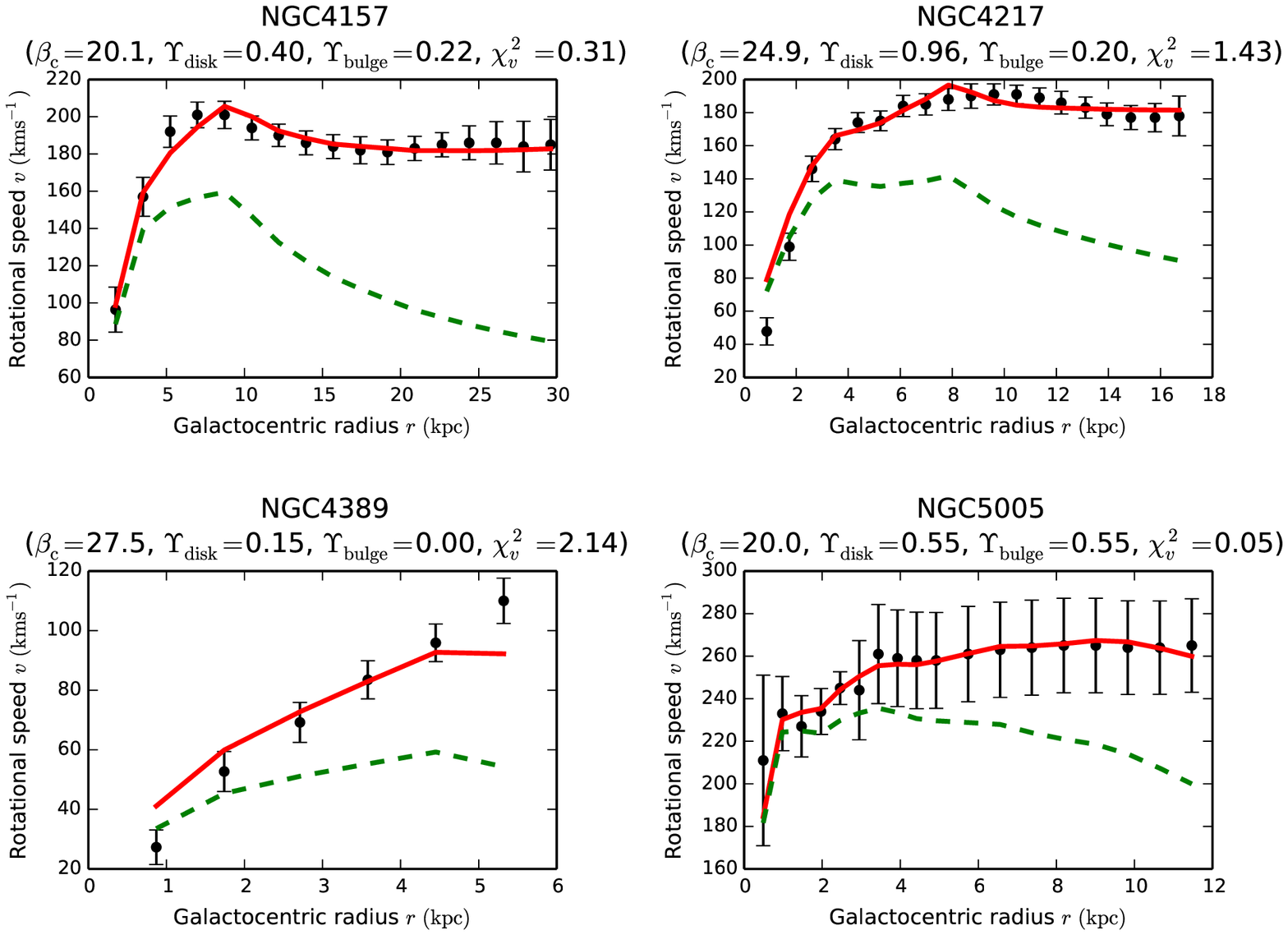}
 \caption{The resultant rotation curve of the NGC 4157, NGC 4217, NGC 4389 and NGC 5005 galaxies. Here, the green dashed line is the rotation curve contributed by the baryonic component with $x=0$. The red solid line is the resultant rotation curve based on our model. The rotation curve data with error bars are taken from \citet{Lelli}.}
\vskip 10mm
\end{figure}

\begin{figure}
\vskip 10mm
 \includegraphics[width=140mm]{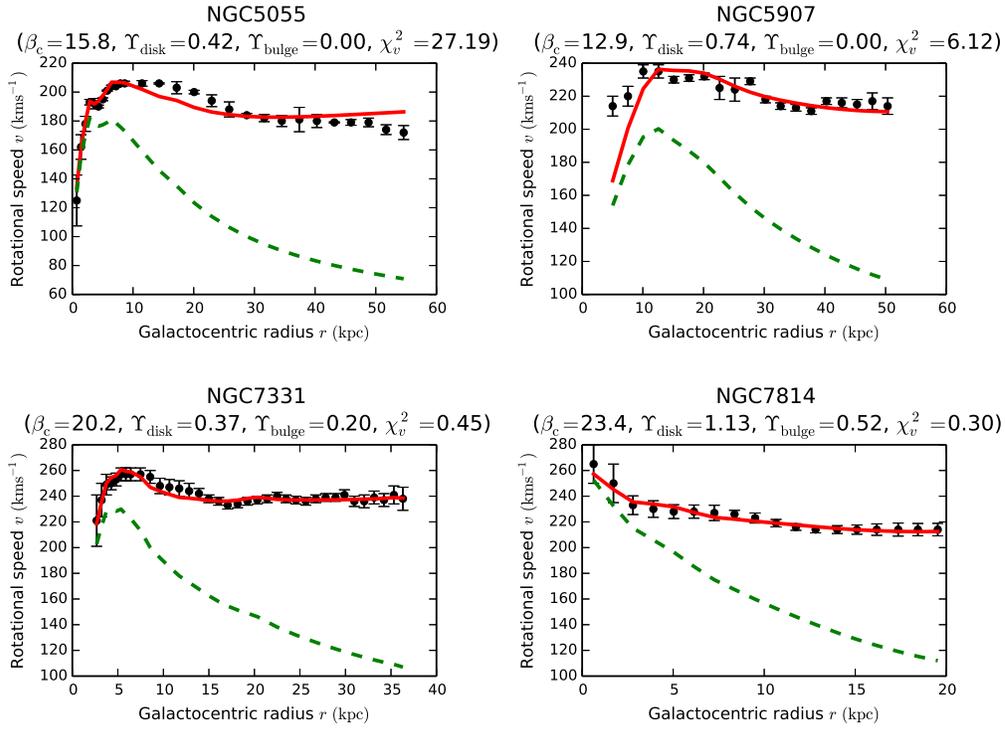}
 \caption{The resultant rotation curve of the NGC 5055, NGC 5907, NGC 7331 and NGC 7814 galaxies. Here, the green dashed line is the rotation curve contributed by the baryonic component with $x=0$. The red solid line is the resultant rotation curve based on our model. The rotation curve data with error bars are taken from \citet{Lelli}.}
\vskip 10mm
\end{figure}

\section{Discussion}
In this article, we test the CGG model by the Milky Way rotation curve and the SPARC data. We show that the CGG model can provide a satisfactory explanation to the missing mass problem in galaxies. The extra term arising from the CGG model can provide enough extra `gravity' to flatten the outer rotation curves without any dark matter. This model can also provide a solution to the disk-halo conspiracy problem. The only free parameter in this model is constrained to $\beta \approx 20$. This value does not violate any observational constraint based on our Solar system.  

However, recent studies with cosmological data of the Integrated Sachs-Wolfe (ISW) effect \citep{Renk} and bounds on gravitational wave speed \citep{Lombriser,Lombriser2,Creminelli,Ezquiaga} seem to rule out the CGG as a cosmological viable theory. In order to match the cosmological constraints (e.g. the cosmic microwave background data), the evolution of the Galileon field must satisfy a so-called tracker solution before the energy density of the Galileon field starts to contribute non-negligibly to the total energy density of our universe \citep{Renk}. The ISW data and the bounds on gravitational wave speed suggest difficulties with the CGG based on the tracker solution. Nevertheless, our focus in this article is not CGG as an alternative model in cosmology without dark energy, but an alternative to dark matter. The Galileon field is in general a free parameter in the CGG theory \citep{Renk}. Here, the field used in this analysis satisfies the Poisson equation instead of the cosmological tracker solution. Besides, although the Vainshtein Radius $r_V$ has a term $\Lambda$, its physical meaning is not related to the dark energy. We only take it with the order of the Hubble constant $H_0$. Therefore, it is possible that CGG as an alternative to dark energy is wrong but CGG as an alternative to dark matter is correct. Dark energy may still exist within the CGG model. Moreover, the cosmological evidence can sometimes be ambiguous. It is too early to conclude that CGG is ruled out by these studies (even as an alternative to dark energy). For example, a recent study shows that the problem of the tracker solution in CGG theory due to the bounds on gravitational wave speed can be avoided by the existence of a quadratic k-essence Lagrangian \citep{Kase}. Also, some recent studies point out that the mass of neutrinos might affect the cosmological ISW data, though it cannot completely rescue the CGG theory \citep{Peirone}.  

On the other hand, some studies point out that the observational data of the Bullet Cluster 1E0657-558 may give some challenge to the alternative theory of gravity, especially for MOND \citep{Takahashi}. The spatial offset of the center of the total mass from the center of the baryonic mass is difficult to be explained with the alternative theory of gravity. Nevertheless, \citet{Brownstein} show that the observed cluster thermal profile gives good agreement with the MOG. The theory of MOG gives some extra terms in Newtonian gravity which demonstrate similar effects with the CGG model. Although we have not applied the CGG model to the Bullet Cluster, we believe that it is not a big problem to the CGG model. In fact, it is also a controversial issue whether the standard $\Lambda$CDM model is consistent with the data of the Bullet Cluster collision \citep{Lee,Lage}. Therefore, the case of the Bullet Cluster should not be treated as a smoking gun to rule out any alternative theory of gravity.

As mentioned above, the null detection of dark matter in direct detection experiments might indicate that the assumption of the existence of particle dark matter is wrong. Therefore, alternative theories of gravity can provide another way to address the missing mass problem in galaxies, galaxy clusters and our entire universe. Not only MOG can achieve this purpose, the CGG model can also achieve the same goal. Further observational tests for dwarf galaxies might be able to differentiate which alternative theories of gravity can provide the best explanation to the missing mass problem.

\begin{acknowledgements}
This work is supported by the Dean's Research Fund from The Education University of Hong Kong (Project No.:SFRS9 2017).
\end{acknowledgements}

\end{document}